\documentclass[twocolumn,preprintnumbers,amsmath,amsfonts,amssymb,floatfix,aps,pra,superscriptaddress]{revtex4}

\usepackage[utf8]{inputenc}
\usepackage{graphicx}
\usepackage{amssymb}
\usepackage{amsmath}
\usepackage{amsfonts}
\usepackage{bm}
\usepackage{color}
\usepackage{hyperref}

\newcommand{\be}{\begin{equation}}
\newcommand{\ee}{\end{equation}}

\begin{document}

\title{Achieving one-dimensionality with attractive fermions}

\author{F. Chevy}%
\email{frederic.chevy@phys.ens.fr}
\affiliation{Laboratoire de Physique de l’\'Ecole normale sup\'erieure, ENS, Universit\'e PSL, CNRS, Sorbonne Universit\'e, Universit\'e Paris Cit\'e, F-75005 Paris, France%\\This line break forced% with \\
}%
%\affiliation{Laboratoire Kastler Brossel, ENS-Universit\'e PSL, CNRS, Sorbonne Universit\'e, Coll\'ege de France.%\\This line break forced% with \\
%}%

\author{G. Orso}%
\email{giuliano.orso@u-paris.fr}
\affiliation{Universit\'e Paris Cit\'e, Laboratoire Mat\'eriaux et Ph\'enom\`enes Quantiques (MPQ), CNRS, F-75013, Paris, France %\\This line break forced% with \\
}%

\begin{abstract}
In this article we discuss the accuracy of effective one-dimensional theories used to describe the behavior of ultracold atomic ensembles confined in quantum wires by a harmonic trap.  We derive within a fully many-body approach the effective Hamiltonian describing this class of systems and we calculate the beyond-mean field corrections to the energy of the ground state arising from virtual transitions towards excited state of the confining potential. We find that, due to the Pauli principle,  effective finite-range corrections are one of magnitude larger than effective three-body interactions.
By comparing to exact solutions of the purely 1D problem, we conclude that a 1D effective theory provides a good description of the ground state of the system for a rather large range of interaction parameters.   
\end{abstract}

\maketitle

\section{Introduction}

Among quantum technologies, quantum simulation aims at finding the properties of complex Hamiltonians by engineering experimental systems whose dynamics can be described as precisely as possible by the problem under study \cite{bloch2008many,Georgescu2014QuantumSimulation}. Within this program,  ultracold atoms have been used in the past decade to solve Bertsch's many-body X-challenge on the structure of strongly correlated quantum patter \cite{zwerger2012BCSBEC}, or to emulate lattice models \cite{Browaeys2020Many-bodyAtoms}. 

In this context the control of the experimental parameters is paramount to the success of the quantum simulation program and in this article we discuss the feasibility of the simulation of low dimensional systems using ultracold systems. Experimentally, low-dimensionality can be achieved by strongly compressing particles along one or two dimensions. When the energy of the particles is lower than the distance between the ground state and the first excited state of the trapping potential, the dynamics is frozen along these directions and we can consider the system as being kinematically 1D or 2D. 

In this article, we focus on one-dimensional fermionic systems. These systems are accessible experimentally using cold atoms and have been used in recent years to study a  large array of phenomena, such as their thermodynamic properties \cite{DeDaniloff2021InWires} or pairing close to confinement induced resonances \cite{moritz2005confinement}. Their phase diagram in the presence of some spin-imbalance \cite{Orso2007Attractive,Hu2007Phase,parish2007quasi,liao2009spin,revelle20161d}, or even in large spin systems \cite{pagano2014one} was explored. Their structure factor has been characterized \cite{yang2018measurement} and spin-charge separation was observed \cite{He2020EmergenceFermions}. More importantly for our purpose here, their properties can be calculated exactly using Bethe Ansatz \cite{guan2013fermi} and these solutions can be used as a reliable benchmarks to quantify deviations from true one-dimensionality in a realistic experimental system. 

The mechanism leading to a low-dimensional regime is however not necessarily valid for strongly correlated systems. As pointed out in earlier works, virtual transitions towards excited states of the confining potential can modify the effective interactions between particles by giving rise to emergent few-body interactions \cite{Mazets08breakdown,tan10relaxation,pricoupenko19three,Goban2018EmergenceClock} and even modify the phase diagram of the system \cite{Fuchs2004ExactlyCrossover,Tokatly2004DiluteGas,Mora2005Four-bodyGas}. This departure from pure one-dimensionality is potentially more pronounced for fermions that are intrinsically stable close to Feshbach resonances.

% and we will explore the possibility of simulating the celebrated Yang-Gaudin Hamiltonian \cite{Gaudin1967UnInteraction,Yang1967SomeInteraction}. in realistic ultracold atom experiment. We will show that this is possible only in relatively weakly interacting systems and that in the strongly correlated regime, the properties of strongly confined systems cannot be captured by a purely 1D effective model. 
By considering first the transverse size of the cloud, we show in Sec. \ref{sec:SizeArgument} that for repulsive interactions of arbitrary strength, the Yang-Gaudin regime can be achieved as long as the density is small enough. By contrast, we show for strongly attractive systems that the occupation of excited states remain finite even for vanishingly small densities and that true one-dimensionality can therefore never be achieved in this regime. In the following sections, we focus on the weakly attractive limit where Yang-Gaudin's limit can be achieved. Using Schrieffer-Wolff's approach \cite{Schrieffer1966RelationHamiltonians}, we derive the many-body effective Hamiltonian describing the low-energy physics of fermions in a quantum wire (Sec. \ref{Sec:EffectiveHamiltonian}). We recover effective three-body interactions found in previous works on quasi-1D few-body physics and we use them to calculate the first beyond-mean-field corrections to the energy of the many-body system. We conclude that even for rather large Fermi energy, the corrections to the Yang-Gaudin Hamiltonian remain small.

\section{The Yang-Gaudin Hamiltonian}

The Yang-Gaudin Hamiltonian \cite{Gaudin1967UnInteraction,Yang1967SomeInteraction} is one of the simplest models introduced in quantum many-body physics. It describes an ensemble of one-dimensional spin 1/2 fermions with contact interactions and is expressed as 

\be
H_{\rm YG}=\sum_{i,\sigma}\frac{p_{\sigma,i}^2}{2m}+g_{1D}\sum_{i,j}\delta(z_{\uparrow,i}-z_{\downarrow,j}).
\ee
Here, $m$ is the mass of the particles, $p_{\sigma, i}$ and $z_{\sigma,i}$ are respectively the momentum and the positions of the $i$th particle carrying a spin $\sigma=\updownarrow$ and $g_{1D}$ is a coupling constant that can be expressed using a so-called 1D-scattering length $a_{1D}$ defined by $g_{1D}=-2\hbar^2/m a_{1D}$.

The ground state of the Yang-Gaudin Hamiltonian can be found analytically using Bethe's Ansatz \cite{guan2013fermi}. 
For attractive interactions and in the absence of spin-imbalance, this exact solution amounts to solving numerically the integral equation for the spectral function
\be
\sigma(\lambda)=\frac{1}{\pi} - \frac{1}{\pi} \int_{-B}^{B} \frac{1}{1+ (\lambda-\lambda^\prime)^2} \sigma(\lambda^\prime) d\lambda^\prime,
\ee 
where $B$ is a positive quantity, which is related to the total particle density $n$ by 
\be
n a_{1D}=4 \int_{-B}^{B} \sigma(\lambda) d\lambda.
\ee 
The ground state energy per unit of length is given by
\be
\frac{E_{\rm YG}}{L}=\frac{4\hbar^2}{m a_{1D}^3}\int_{-B}^{B} \left (2\lambda^2-\frac{1}{2} \right ) \sigma(\lambda) d\lambda.
\label{Eq:BetheAnsatz}
\ee

On a  dimensional ground, the solutions of these equations are characterized by a single dimensionless parameter $\gamma=-2/na_{1D}$ that compares the kinetic and interaction energies of the system. $\gamma=0$ corresponds to a non-interacting system while $|\gamma|\rightarrow\infty$ corresponds to a strongly interacting regime. For negative $\gamma$, this corresponds to a ``fermionized" gas of bosonic dimers with a binding energy  $\epsilon^b_{1D}=\hbar^2/ma_{1D}^2$.

In this article we focus on weakly attractive systems. This limit is singular using Bethe's Ansatz approach, leading to contradictory claims about the behaviour of beyond mean-field corrections \cite{guan2013fermi,Iida2007ExactPolarization,Krivnov1975One-dimensionalElectrons}. We will then rather proceed with a direct perturbative expansion using Rayleigh-Schr\"odinger's formalism, that will also be more easily amenable to the study of quasi-1D  systems. Using second order perturbation theory, we readily see that the energy of an ensemble of spin 1/2 fermions is given by 

\be
\begin{split}
E_{\rm YG}=&E_0+g_{1D}\frac{nN}{4}\\&+\frac{g_{1D}^2}{L^2}\sum_{\substack{|q_\sigma|<k_{F}\\|p_\sigma|>k_F\\p_\uparrow+p_\downarrow=q_\uparrow+q_\downarrow}}\frac{m}{\hbar^2(q^2-p^2)}+....,
\end{split}
\label{Eq:Perturbation1D}
\ee
where $E_0=NE_F/3$ is the energy of the non-interacting system, $k_F=n\pi/2$ is the Fermi wave vector, $p_{\sigma}$ and $q_{\sigma}$ are the momenta of the particles and holes of spin $\sigma$ created by the interaction and $q=(q_\uparrow-q_\downarrow)/2$ and $p=(p_\uparrow-p_\downarrow)/2$ are their relative momenta. The sum appearing in the first beyond-mean-field contribution can be calculated analytically and we obtain

\be
E_{\rm YG}\simeq\frac{NE_F}{3}\left[1+\frac{6\gamma}{\pi^2}-\frac{\gamma^2}{\pi^2}\right].
\label{Eq:BeyondMF}
\ee
The above result is analytical in density and is identical to the one previously reported in \cite{Marino2019ExactModels} using a direct asymptotic expansion of the Bethe-Ansatz solution. In particular, we do not find the logarithmic contribution initially predicted by Krivnov and Ovchinnikov \cite{Krivnov1975One-dimensionalElectrons}. 

\begin{figure}
    \centering
    \includegraphics[width=\columnwidth]{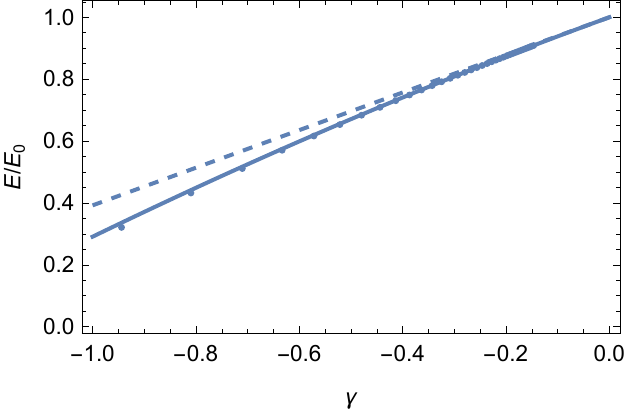}
    \caption{Energy of an attractive quasi-1D Fermi gas in the weakly attractive limit. The dots correspond to the exact solution the true 1D system using Bethe Ansatz's solution (Eq. \ref{Eq:BetheAnsatz}). The dashed line is the asymptotic solution incorporating the first beyond mean-field correction. We see that the agreement extends up to $|\gamma|\lesssim 1$. }
    \label{fig:FigGraph}
\end{figure}

In Fig. \ref{fig:FigGraph}, we compare the second-order expansion (\ref{Eq:BeyondMF}), (solid line), to the exact result obtained using the Bethe-Ansatz (dot symbols). We see that they both agree in a rather broad parameter regime $|\gamma|\lesssim 1$.

\section{Quantum-simulation of Yang-Gaudin's Hamiltonian}
\label{sec:SizeArgument}
Experimentally, effective low dimensional physics is achieved by confining very strongly an ensemble of particles in one, two or three dimensions. When the typical single particle energies (ie chemical potential, temperature) are much smaller than the energy between the ground state and the first excited state of the trapping potential, it is usually assumed that the dynamics along the confined directions is frozen and that the system becomes effectively one- or two-dimensional, depending on the number of frozen directions \cite{cazalilla2011one}. This mechanism is used in condensed matter physics, for instance at the junction between N- and P-doped semi-conductors in order to realize 2D-electron gases. In cold atoms, 1D and 2D vapours have been realized using optical trapping or magnetic trapping at the surface of atom chips.     

In the ultracold regime, the de Broglie wave-length of the atoms is much larger than the range of the interatomic potential. As a consequence, interactions in free space can be accurately modeled using a zero-range contact potential characterized by a 3D scattering length $a_{3D}$. One would then naively expect that the low-energy physics in confined geometry would be described by a YG Hamiltonian, with a coupling constant $g_{1D}$ depending on $a_{3D}$ and the trap parameters. 

Let's consider a quasi-1D system described by the Hamiltonian

\be
H=\sum_i \frac{p_{\sigma, i}^2}{2m}+\sum_{i,\sigma}\frac{m\omega_\perp^2}{2}(x_{\sigma,i}^2+y_{\sigma,i}^2)+\sum_{i,j}V(\bm r_{\uparrow,i}-\bm r_{\downarrow,j})
\ee
where $V$ is the 3D interparticle interaction potential. Let's note $E$ the energy of the ground-state. Assuming that the dynamics is indeed frozen in the $(x,y)$ plane, $E$ can be written as
\be
E=N\hbar\omega_\perp+E_{\rm YG},
\ee
where $E_{\rm YG}$ is the ground state energy of the YG model with a 1D scattering length expressed in terms of $a_{ 3D}$ and $\ell_\perp=\sqrt{\hbar/m\omega_\perp}$. In the case of the two-body problem, the value of $a_{1D}$ was calculated in \cite{olshanii1998atomic} and we have in this case:
\be
a_{1D}=-\frac{\ell_\perp}{2}\left(\frac{\ell_\perp}{a_{3D}}+\zeta(1/2)\right),
\ee
where $\zeta$  is Riemann's Zeta function. The transverse size of the cloud can be deduced from $E$ using Hellman-Feynman's theorem. We have indeed:
\be
\langle\rho^2\rangle=\frac{2}{Nm}\frac{\partial E}{\partial\omega_\perp^2},
\ee
where $\rho^2=x^2+y^2$.
Since $E_{\rm YG}$ depends implicitly on $\omega_\perp$ through $a_{1D}$, we have

\be
\langle\rho^2\rangle=\ell_\perp^2+\frac{1}{N}\frac{\partial a_{1D}}{\partial \ell_{\perp}}\frac{\partial E_{\rm YG}}{\partial a_{1D}}
\ee

The first term corresponds to the size of the ground state and in the second term we recognize the so-called Tan-contact parameter \cite{tan2008large,Olshanii2003Short-DistanceGases}. We can consider the system as effectively 1D if $\langle\rho^2\rangle$ stays close to $\ell_\perp^2$ (note that, paradoxically, compressing the transverse size below $\ell_\perp^2$ does not improve one-dimensionality. Indeed, due to Heisenberg's uncertainty relations, this implies that additional transverse energy is stored as kinetic energy). 

In the weakly interacting limit, we can use Eq. (\ref{Eq:BeyondMF}) to evaluate  $E_{\rm YG}$ within the mean-field approximation. In this case, the transverse size of the system is given by 

\be
\langle\rho^2\rangle\simeq \ell_\perp^2\left(1-\frac{\ell_\perp^2}{a_{1D}^ 2}\frac{\partial a_{1D}}{\partial\ell_\perp} n\ell_\perp\right).
\ee
Interestingly, we see that when the density is small the correction to the non-interacting case can also be made arbitrary small thanks to the $n\ell_\perp$ contribution. In other words, In the mean-field regime, the transverse radius is consistent with a frozen motion in the transverse direction as long as the Fermi energy $E_F=\hbar^2 (n\pi)^2/8m$ is sufficiently small compared to $\hbar\omega_\perp$ as expected from the naive interpretation of the quasi-1D regime. 

Let's now consider the strongly attractive limit (corresponding to a large $\gamma$). In this case, the 1D system is a gas of bosonic dimers and $E_{\rm YG}$ is dominated by their binding energies, ie 

\be
E_{\rm YG}\simeq-\frac{N\hbar^2}{2ma_{1D}^2}.
\ee
We then obtain 
\be
\langle\rho^2\rangle=\ell_\perp^2\left(1-\frac{\ell_\perp^3}{2a_{1D}^ 3}\frac{\partial a_{1D}}{\partial\ell_\perp}\right).
\ee
We see that in this case, the correction becomes density independent and as a consequence, a finite amount of energy remains stored in the transverse degrees of freedom, even when $E_F\ll\hbar\omega_\perp$. Close to the confinement-induced resonance where $a_{1D}$ vanishes, the correction even diverges. We conclude from this argument that the quasi-1D approximation is no longer valid when the system approaches the resonant  $a_{3D}/\ell_\perp\rightarrow\infty$, and beyond.   

The origin of this breakdown was outlined in past references \cite{Fuchs2004ExactlyCrossover,Tokatly2004DiluteGas,Mora2005Four-bodyGas}: when entering the strongly attractive regime, the molecule binding energy becomes larger than $\hbar\omega_\perp$. As a consequence their internal structure is no longer affected by the transverse confining potential and becomes identical to that of molecules in free space. This phenomenon is more dramatically illustrated by the fact that in a quasi-1D geometry, there is a bound state for any value on the 3D scattering length, even when $a_{1D}$ is negative,  which contradicts the fact that for a purely 1D system,  bound state exist only for positive $a_{1D}$.

\section{Effective Hamiltonian for quasi-1D fermions}
\label{Sec:EffectiveHamiltonian}
Since Yang-Gaudin's Hamiltonian ceases to apply for strong interactions, we focus on the weakly interacting regime and we explore its accuracy within a perturbative approach. According to Fig. (\ref{fig:FigGraph}), a calculation to second order in perturbation theory should be valid up to $\gamma\simeq 1$. Interactions give rise to both ``intraband" processes, where all particles stay in the transverse ground state of the trapping potential, and ``interband" collisions where one or more particles are virtually excited towards an excited state of the confining potential. It is well known that these virtual processes give rise to effective few-body interactions \cite{Goban2018EmergenceClock} that we calculate in a many-body context using the Schrieffer-Wolff's (SW) approach \cite{Schrieffer1966RelationHamiltonians}. 
The many-body Hamiltonian describing an ensemble of spin-1/2 fermionic atoms confined by a transverse external potential is given by
\be\label{Ham}
H=H_0+H_{\rm int},
\ee
In this expression, $H_0$ is the free particle Hamiltonian. In a second-quantized form, we write it as
\be
H_0=\sum_{\alpha,\sigma}\epsilon(k,\alpha)a_{\sigma k\alpha}^\dagger a_{\sigma k\alpha}
\ee
where $\sigma$ is the spin index, $k$ is the momentum in the $z$ direction and $\alpha=(n,m_z)$ labels the transverse state: $n$ is  the quantum number associated with the energy of the transverse motion and $\hbar m_z$ is the angular momentum along $z$. For a harmonic confinement,  $m_z\in\{-n,-n+2,...n-2,n\}$ and $\varepsilon(k,\alpha)=\hbar^2 k^2/2m+\hbar\omega_\perp(n+1)$. 

$H_{\rm int}$ describes the two-body interactions and is given by 

\be
H_{\rm int}=\frac{g_{3D}}{L}\sum_{\substack{k_1+k_2=\\k_3+k_4\\(\alpha_1\alpha_2\alpha_3\alpha_4)}}\chi_{\alpha_1\alpha_2\alpha_3\alpha_4}a^\dagger_{\uparrow k_1 \alpha_1}a_{\uparrow k_2 \alpha_2}a^\dagger_{\downarrow k_3 \alpha_3}a_{\downarrow k_4 \alpha_4},
\ee
where $g_{3D}=4\pi\hbar^2 a_{3D}/m$ is the bare coupling constant and the matrix elements $\chi_{\alpha_1\alpha_2\alpha_3\alpha_4}$ are defined as 
%\begin{eqnarray}\chi_{\alpha_1\alpha_2\alpha_3\alpha_4}&=&\langle \psi_{\alpha_1}\psi_{\alpha_2}|\delta(\bm\rho_1-\bm\rho_2)|\psi_{\alpha_3}\psi_{\alpha_4}\rangle\\
%&=&\int d^2\bm\rho\psi_{\alpha_1}(\bm\rho)^*\psi_{\alpha_2}(\bm\rho)^*\psi_{\alpha_3}(\bm\rho)\psi_{\alpha_4}(\bm\rho),
%\end{eqnarray}
%\toadd
{\begin{equation}
\chi_{\alpha_1\alpha_2\alpha_3\alpha_4}=
\int d^2\bm\rho\psi_{\alpha_1}(\bm\rho)^*\psi_{\alpha_2}(\bm\rho)\psi_{\alpha_3}^*(\bm\rho)\psi_{\alpha_4}(\bm\rho),    
\end{equation}
with $\psi_\alpha(\bm\rho)$ being the wave function associated with the eigenstate $\alpha$ of the transverse motion. }

To implement the Schrieffer-Wolff scheme, we write $H_{\rm int}=H_1+H_2$, where $H_1$ and $H_2$ correspond  to intra- and interband processes, ie to contributions of  $(\alpha_1,\alpha_2,\alpha_3,\alpha_4)=(0,0,0,0)$ and $(\alpha_1,\alpha_2,\alpha_3,\alpha_4)\not =(0,0,0,0)$ respectively.

Since $H_2$ is responsible to virtual interband transitions, we treat it perturbatively within SW approach. Lets define $\tilde H=H_0+H_2$. In the  SW scheme we look for a canonical transformation 
$\tilde H^\prime =\exp(S) \tilde H \exp(-S)$ with generator $S$ such that $\tilde H^\prime$ becomes diagonal to first order in $H_2$. This amounts to requiring
\begin{equation}\label{SWa}
[H_0,S]=H_2,    
\end{equation}
yielding
\begin{equation}
\tilde H^\prime =H_0+\frac{1}{2} [S,H_2]+\mathcal{O}(g_{3D}^3),    
\end{equation}
showing that the correction to the Hamiltonian $\tilde H$ is  
quadratic in the coupling constant as $S$ is linear in $g_{3D}$. 
The total Hamiltonian in the rotated basis is then given by
\begin{equation}
H^\prime = e^S H e^{-S} =\tilde H^\prime +  e^S H_1 e^{-S}=H_{\rm eff}+   \mathcal{O}(g_{3D}^3), \label{SWc}
\end{equation}
where 
\begin{equation}\label{Heff}
H_{\rm eff}=H_0+H_1+\frac{1}{2} [S,H_2]+[S,H_1]
\end{equation}
is the perturbative effective Hamiltonian of the system that we are looking for. Notice that for $S=0$, the effective Hamiltonian reduces to the Yang-Gaudin model,  since $H_0+H_1=H_{\rm YG}$.  

In order to write $H_{\rm eff}$ explicitly in second quantization form, we need to find a representation of the generator $S$ in terms of the fermionic field operators. Since $H_0$ and $H_2$ are, respectively, quadratic and quartic in those fields, it is easy to see that Eq.~(\ref{SWa}) can only be satisfied if $S$ is a quartic operator of the form
\begin{equation}\label{SwS}
S= \!\!\frac{g_{3D}}{L} \!\!\! \sum_{\substack{k_1, k_2, k_3, k_4\\(\alpha_1,\alpha_2,\alpha_3,\alpha_4) \not =\\(0,0,0,0)}} \!\!\! f(\{k_i,\alpha_i\})a_{\uparrow k_1 \alpha_1}^\dagger  a_{\downarrow k_3 \alpha_3}^\dagger a_{\downarrow k_4 \alpha_4} a_{\uparrow k_2 \alpha_2},   
\end{equation}
where $f(\{k_i,\alpha_i\})$ is an unknown function of the four axial momenta $k_i$ and the four discrete indices $\alpha_i$ for the motion along the transverse directions.
Notice that the term with all $\alpha_i=0$ in Eq.(\ref{SwS} does not mix with the excited states of the confining potential and therefore cannot contribute to the generator $S$ (otherwise $[H_0,S]$ would also contain a similar term, which is instead absent in $H_2$, thus violating Eq.(\ref{SWa})). 

In order to determine the function $f$, we substitute Eq.(\ref{SwS}) into Eq.(\ref{SWa}) and use the anticommutation relations $\{a_{\sigma k \alpha}^\dagger, a_{\sigma^\prime p \alpha^\prime}\}=\delta_{\sigma\sigma^\prime} \delta_{kp}\delta_{\alpha \alpha^\prime}$, $\{a_{\sigma k \alpha}, a_{\sigma^\prime p \alpha^\prime}\}=0$. This yields
%The details of the derivation of $f$ and of the effective Hamiltonian are given in Apprendix \ref{Appendix1}. The function $f$ appearing in $S$ is given by
\begin{equation}\label{f}
f(\{k_i,\alpha_i\})= \frac{\chi_{\alpha_1 \alpha_2 \alpha_3 \alpha_4}\delta_{k_1+k_3,k_2+k_4}}{\epsilon(k_1,\alpha_1)+\epsilon(k_3,\alpha_3)-
\epsilon(k_2,\alpha_2)-\epsilon(k_4,\alpha_4)}.   
\end{equation}

%\toadd
The effective Hamiltonian can then be calculated by substituting Eq.s (\ref{SwS}) and (\ref{f}) in Eq.~(\ref{Heff}), and by evaluating the two commutators. The details of the derivation are given in Appendix \ref{Appendix1}. In particular, since we are interested on the 1D effective theory describing atoms in the ground state of the harmonic oscillator, we consider only creation and annihilation operators with $\alpha=0$. This implies that $[S,H_1]$ does not contribute to the effective model 
%and that the contribution of the commutator $[S,H_2]$ can be decomposed into a 2-body part (containing only four creation and annihilation operators) 
%and a three body sector (containing products of six operators). We thus write $[S,H_2]/2=H'_2+H''_2$ with
while the contribution from the commutator $[S,H_2]$ can be recast as a sum of two parts,
$[S,H_2]/2=H'_2+H''_2$. The first part describes two-body interactions  (containing only four creation and annihilation operators) and is given by
\be
H'_2=\frac{g_{3D}^2}{2L^2}\sum_{\substack{k_1+k_3\\=\\p_2+p_4}} 
      (F(k_1,k_3)+F(p_2,p_4))\;
    a_{\uparrow k_1 0}^\dagger  a_{\downarrow k_3 0}^\dagger a_{\downarrow p_4 0} a_{\uparrow p_2 0},
\ee
 with 
\be
F(k_1,k_3)=\sum_{\substack{k_2,k_4\\\alpha_2\alpha_4}}f(\{k_i,\alpha_i\})\chi_{\alpha_2 0\alpha_4 0}.
\ee
The second part corresponds to emergent three-body interactions (containing products of six operators) and can be written as
\be
\begin{split}
&H''_2=\frac{g_{3D}^2}{2L^2}\sum_{\substack{k_1+k_3+p_3\\=\\p_2+p_4+k_4}} 
 \!\!\!\Big[\left(G(k_1,k_3,k_4)+G(p_2,p_3,p_4)\right)\times\\
 &a_{\uparrow k_1 0}^\dagger   a_{\downarrow k_3 0}^\dagger a_{\downarrow p_3 0}^\dagger a_{\downarrow p_4 0} a_{\downarrow k_4 0} a_{\uparrow p_2 0} + (\uparrow \leftrightarrow \downarrow)\Big],
\end{split}
\label{H2''}
\ee
with
\be
G(k_1,k_3,k_4)=\sum_{k_2 \alpha_2}f(\{k_i,\alpha_i\})\chi_{\alpha_2 000}.
\label{Eq:G}
\ee

\section{Effective two-body interactions}
\label{Sec:2body}
By construction, $H''_2$ in Eq. (\ref{H2''}) does not contribute to the two-body sector. 
%The system can be described by an effective two-body interaction that we reorganize as $H_1+H'_2=U'+U''$ with
We reorganize the effective two-body interaction at low energy as $H_1+H'_2=U'+U''$ with

\be
U'=\frac{\tilde g_{1D}}{L}\sum_{k_1+k_2=p_1+p_2}
    a_{\uparrow k_1 0}^\dagger  a_{\downarrow k_2 0}^\dagger a_{\downarrow p_2 0} a_{\uparrow p_1 0},
\ee
where the effective 1D coupling constant is given by
\be
\begin{split}
\tilde g_{1D}&=g_{3D}\chi_{0000}\\&-\frac{g_{3D}^2}{L}\sum_{\substack{k_2+k_4=0\\ \alpha_1\alpha_2}}\frac{\chi_{0\alpha_10\alpha_2}^2}{\frac{\hbar^2k_2^2}{2m}+\frac{\hbar^2k_4^2}{2m}+\hbar\omega_\perp(n_{\alpha_2}+n_{\alpha_4})}.
\end{split}
\label{Eq:Effectiveg1D}
\ee
Strictly speaking, this sum is divergent (see appendix \ref{Sec:A2}). This is a consequence of the zero-range potential approximation that is notoriously known to be singular. The divergence can be cured by the introduction of a UV-cutoff and by considering $g_{3D}$ as a running  coupling constant that vanishes when the cutoff goes to zero. If properly performed, this regularization procedure recovers the exact result derived in \cite{olshanii1998atomic}. Here, we wil take $\tilde g_{1D}$ as given and we will use it to renormalize all diverging quantities appearing in forthcoming calculations. 

The remaining part of the two body interaction is given by
\be
\begin{split}
U''=\frac{g_{3D}^2}{2L^2}\sum_{k_1+k_3=p_2+p_4}& 
      (F(k_1,k_3)+F(p_2,p_4)-2F(0,0))\\
      & \times   a_{\uparrow k_1 0}^\dagger  a_{\downarrow k_3 0}^\dagger a_{\downarrow p_4 0} a_{\uparrow p_2 0}
\end{split}
\ee
and describes finite-range/momentum-dependent corrections to the two-body interactions. Note that contrary to $\tilde H_1'$, this term has a well-defined UV limit. 

\section{Effective three-body interaction}

Let's consider now the low-energy behaviour of the three-body interaction term $H_2''$. When taking $k_{1,3,4}=p_{2,3,4}=0$ in Eq. (\ref{H2''}), the exchange of $k_3$ and $p_3$ on the one hand, and $k_4$ and $p_4$ on the other hand lead to a cancellation of $H''_2$. This is a departure from previous results on bosonic systems \cite{Mazets08breakdown,tan10relaxation,pricoupenko19three} where the three-body coupling constant is momentum independent. Indeed, let's consider a generic three-body interaction 

\be
\begin{split}
&U_{3\rm b}=\sum_{\substack{k_1+k_3+p_3\\=\\p_2+p_4+k_4}} 
\!\!\!\Big[K(\{k_i,p_j\})\times\\
 &a_{\uparrow k_1 0}^\dagger   a_{\downarrow k_3 0}^\dagger a_{\downarrow p_3 0}^\dagger a_{\downarrow p_4 0} a_{\downarrow k_4 0} a_{\uparrow p_2 0} + (\uparrow \leftrightarrow \downarrow)\Big]
\end{split}
\ee
If we expand the function $K$ to second order in momentum, we readily see that the lowest order term yielding a nonzero contribution because of fermionic exchange is 
\be
K\propto (k_3-p_3)(k_4-p_4).
\ee
This result can also be recovered directly by calculating the low momentum asymptotic behaviour of Eq. (\ref{Eq:G}). For this, we  take $k_1=p_2=0$ in $G$ and then expand it to first order in $k_{3,4}$ and $p_{3,4}$. We then obtain

\be
\begin{split}
&H''_2=-\frac{\hbar^2a_{3D}^2}{4mL^2}{\rm Li}_{1/2}(1/4)\sum_{\substack{k_1+k_3+p_3\\=\\p_2+p_4+k_4}} 
\!\!\!\Big[(k_3-p_3)(k_4-p_4)\times\\
 &a_{\uparrow k_1 0}^\dagger   a_{\downarrow k_3 0}^\dagger a_{\downarrow p_3 0}^\dagger a_{\downarrow p_4 0} a_{\downarrow k_4 0} a_{\uparrow p_2 0} + (\uparrow \leftrightarrow \downarrow)\Big]
\end{split}
\ee
where ${\rm Li}_s(z)=\sum_{n=1}^\infty z^n/n^s$ is the polylog function and where with have used Appendix \ref{Sec:A2} to calculate the sum over $\alpha_2$.

\section{Ground state energy}
Let us now calculate the interaction-induced correction to the ground state energy of the effective Hamiltonian (\ref{Heff}) and verify that we recover the same result by applying perturbation theory to the original Hamiltonian $H$. 
 The noninteracting ground state of the system is given by the product of the Fermi seas of the two spin components, 
 $| FS \rangle=|FS_\uparrow\rangle |FS_\downarrow\rangle$, where
 \begin{equation}
 |FS_\sigma\rangle   =\prod_{|k|<k_{F\sigma}} a_{\sigma k 0}^\dagger|\rangle,
 \end{equation}
where $|\rangle$ represents the vacuum state and $k_{F\sigma}$ are the Fermi momentum of the spin component $\sigma=\uparrow, \downarrow$.

Since the two-body and three-body terms of the effective Hamiltonian arising  from the commutator $[H,S_2]$ are proportional to $g_{3D}^2$, we treat them within \emph{first} order perturbation theory.

Making use of the identities 
\begin{equation}
\begin{split}
   & \langle FS_\sigma | a_{\sigma k_1 0}^\dagger a_{\sigma p_2 0} |FS_\sigma\rangle =
    \Theta(k_{F\sigma}-|k_1|) \delta_{k_1 p_2}   \\
   & \langle FS_\sigma | a_{\sigma k_3 0}^\dagger  a_{\sigma p_3 0}^\dagger
    a_{\sigma p_4 0} a_{\sigma k_4 0}
    |FS_\sigma\rangle =  \Theta(k_{F\sigma}-|k_4|)   \\
   & \quad \quad \times \Theta(k_{F\sigma}-|p_4|)  (\delta_{p_3 p_4}\delta_{k_3 k_4}- \delta_{p_3 k_4}\delta_{k_3 p_4}),
\end{split} 
\label{Eq:comm}
\end{equation}    
we obtain

\begin{equation}
%\label{aa}
\begin{split}
 & \!\!\!\!\!\! \!\!\!\langle FS|\frac{1}{2} [S,H_2]| FS \rangle =  \frac{g_{3D}^2}{L^2} \Bigg[\; \sum_{\substack{|k_1|<k_{F\uparrow} \\ |k_3|<k_{F\downarrow}}} 
 F(k_1,k_3)  \\
  &\!\!\!\!\!\!\!\!-\sum_{\substack{|k_1|<k_{F\uparrow} \\|k_3|,|k_4|<k_{F\downarrow}}}
  %-\sum_{\substack{|k_1|<k_{F\uparrow} \\ |k_3|,|k_4|<k_{F\downarrow}}} 
  \!\!G(k_1,k_3,k_4)
  -\sum_{\substack{|k_1|,|k_2|<k_{F\uparrow} \\ |k_3|<k_{F\downarrow}}}
  \!\!G(k_1,k_3,k_2) \\
 &\!\!\!\!\!\!\!+\!\!\!\sum_{\substack{|k_1|<k_{F\uparrow} \\|k_3|,|p_4|<k_{F\downarrow}}} \!\!G(k_1,k_3,k_3)
  + \!\!\!\!\!\!\sum_{\substack{|k_1| |p_1|<k_{F\uparrow}\\|k_3|<k_{F\downarrow}}} \!\!\!\! G(k_1,k_3,k_1)\Bigg].  
\end{split} 
\label{Eq:pt}
\end{equation}
Notice that  the rhs of  Eq. (\ref{Eq:pt}) contains terms, where two of the 
arguments of the function $G$ become identical. From Eq.(\ref{Eq:G}) we see 
that in this case the function reduces to a constant 
\begin{equation}
  G(k_1,k_3,k_3)=\sum_{m_2 \not =0} -\frac{|\chi_{0 m_2 00}|^2}{m_2 \hbar \omega}.  
\end{equation}

Next, we turn to the correction $\delta E$ to the ground state energy due to $H_1$. Since this term is linear in $g_{3D}$,  we will use \emph{second} order perturbation theory
\be\label{Eq:2orderpert}
\delta E = \langle FS| H_1 | FS\rangle+ \sum_n \frac{ |\langle n| H_1 | FS \rangle|^2}{E_{FS}-E_n},
\ee
where $E_{FS}$ is the energy of the Fermi sea  and $|n\rangle$ is a generic excited eigenstate of $H_0$, which is coupled to 
the ground state by $H_1$. From Eq.(\ref{Eq:comm}) we find that the correction linear in $g_{3D}$ is given by 
\be
\langle FS| H_1 | FS\rangle=\frac{g_{3D}}{L}\chi_{0000}\sum_{\substack{|k_1|<k_{F\uparrow} \\|p_3|<k_{F\downarrow}}} =\frac{g_{3D}}{L}\chi_{0000}N_\uparrow N_\downarrow,
\ee
which corresponds to the mean field. 

%Let's now turn to the beyon mean-field correction. To calculate it, we note that 
The relevant excited states in the rhs of Eq. (\ref{Eq:2orderpert}) correspond to particle-hole excitations $|n\rangle=a_{\uparrow k_1 0}^\dagger a_{\downarrow k_3 0}^\dagger a_{\downarrow k_4 0} a_{\uparrow k_2 0}|FS\rangle$, with the initial states $k_2, k_4$ being inside the respective Fermi surfaces, $|k_2|<k_{F\uparrow},|k_4|<k_{F\downarrow}$, while the final states $k_1, k_3$ are scattered outside them, $|k_1|>k_{F\uparrow},|k_3|>k_{F\downarrow}$. 
From Eq.(\ref{Eq:comm})
we then find 
\be
\langle n| H_1 | FS \rangle=\frac{g_{3D}}{L}\chi_{0000} \delta_{k_1+k_3,k_2+k_4}.
\ee
The change in the kinetic energy of the system brought by the excitation, appearing in the second order correction in Eq.(\ref{Eq:2orderpert}), is given by 
\be\label{den}
E_n-E_{FS}=\frac{\hbar ^2 k_1^2}{2m}+\frac{\hbar ^2 k_3^2}{2m}-\frac{\hbar ^2 k_2^2}{2m} -\frac{\hbar ^2 k_4^2}{2m}.
%\epsilon (k_1,0)+\epsilon (k_3,0)-\epsilon (k_2,0)-\epsilon (k_4,0).
\ee

Finally, we write down the perturbative expansion $E=E_{FS}+\langle FS|[S,H_2]/2| FS \rangle +\delta E$ for the ground state energy for a system of equal spin populations, $N_\uparrow=N_\downarrow=N/2$, corresponding to $k_{F\uparrow}=k_{F\downarrow}=k_F/2$.  
Making use of Eq.s (\ref{defF})-(\ref{defG}), together with Eq.s (\ref{Eq:pt})-(\ref{den}), we find

\begin{widetext}
\be
\begin{split}
& E=E_{FS}+\frac{g_{3D}}{4L} \chi_{0000} N^2 
-\frac{g_{3D}^2}{L^2} \sum_{\substack{|k_1|,|k_3|>k_F \\|k_2|,|k_4|<k_F}} 
\frac{\chi_{0000}^2 \delta_{k_1+k_3,k_2+k_4}}{\frac{\hbar ^2 k_1^2}{2m}+\frac{\hbar ^2 k_3^2}{2m}-\frac{\hbar ^2 k_2^2}{2m} -\frac{\hbar ^2 k_4^2}{2m}} \\
& + \frac{g_{3D}^2}{L^2} \sum_{|k_1|,|k_3|<k_F} \sum_{\substack{k_2,k_4 \\ (n_{\alpha_2},n_{\alpha_4}) \not = (0,0)}} \frac{|\chi_{0 \alpha_2 0 \alpha_4}|^2 \delta_{k_1+k_3,k_2+k_4}}{\frac{\hbar ^2 k_1^2}{2m}+\frac{\hbar ^2 k_3^2}{2m}-\frac{\hbar ^2 k_2^2}{2m} -\frac{\hbar ^2 k_4^2}{2m}-(n_{\alpha_2}+n_{\alpha_4})\hbar \omega}\\
&-2 \frac{g_{3D}^2}{L^2} \sum_{\substack{|k_1|<k_{F} \\|k_3|,|k_4|<k_{F}}} \sum_{k_2, n_{\alpha_2} \not =0} \frac{|\chi_{0m_200}|^2 \delta_{k_1+k_3,k_2+k_4}}{\frac{\hbar ^2 k_1^2}{2m}+\frac{\hbar ^2 k_3^2}{2m}-\frac{\hbar ^2 k_2^2}{2m} -\frac{\hbar ^2 k_4^2}{2m}-n_{\alpha_2}\hbar \omega} - \frac{g_{3D}^2N^3}{4L^2} 
\sum_{n_{\alpha_2}\not =0} \frac{|\chi_{0m_200}|^2}{n_{\alpha_2} \hbar \omega} 
\end{split}
\label{Eq:gdstate}
\ee
\end{widetext}

The second sum appearing in the rhs of Eq. (\ref{Eq:gdstate}) is divergent. Following the prescription outlined in Sec.  \ref{Sec:2body}, this singularity can be cured by noting that its structure is similar to the one of the divergent term of the two-body effective coupling constant $\tilde g_{1D}$.  Indeed, the first three terms of the energy are similar to Eq. (\ref{Eq:Perturbation1D}) for a purely 1D system 
 with a coupling constant constant $g_{1D}=g_{3D}\chi_{0000}/L$. We can recover the same energy, but with the true coupling constant $\tilde g_{1D}$  by adding and subtracting the missing divergent term. We then have

\begin{widetext}
\be
\begin{split}
E=E_{\rm YG}(\tilde g_{1D})&+\frac{g_{3D}^2}{L^2}  \sum_{\substack{|k_1|,|k_3|<k_F\\k_2+k_4=k_1+k_3 \\ (\alpha_2,\alpha_4) \not = (0,0)}} \left[\frac{|\chi_{0\alpha_20\alpha_4}|^2 }{\frac{\hbar ^2 k_{13}}{m}-\frac{\hbar ^2 k_{24}^2}{m} -(n_{\alpha_2}+n_{\alpha_4})\hbar \omega}+\frac{|\chi_{0\alpha_20\alpha_4}|^2 }{\frac{\hbar ^2 k_{24}^2}{m} +(n_{\alpha_2}+n_{\alpha_4})\hbar \omega}\right] \\
&-2 \frac{g_{3D}^2}{L^2} \sum_{\substack{|k_1|<k_{F} \\|k_3|,|k_4|<k_{F}\\k_2+k_4=k_1+k_3\\ \alpha_2 \not =0}}\left[ \frac{|\chi_{0\alpha_200}|^2 }{\frac{\hbar ^2 k_{13}^2}{m}-\frac{\hbar ^2 k_{24}^2}{m} -n_{\alpha_2}\hbar \omega} + \frac{|\chi_{0\alpha_200}|^2}{n_{\alpha_2} \hbar \omega} \right]
\end{split}
\label{Eq:gdstate2}
\ee
\end{widetext}
where $E_{\rm YG}(\tilde g_{1D})$ is the second order expansion of the energy of a 1D system with a coupling constant $\tilde g_{1D}$ given by Eq. (\ref{Eq:Effectiveg1D}) and $k_{ij}=(k_i-k_j)/2$ is the relative momentum of the pair. 

\begin{figure}
    \centering    \includegraphics[width=\columnwidth]{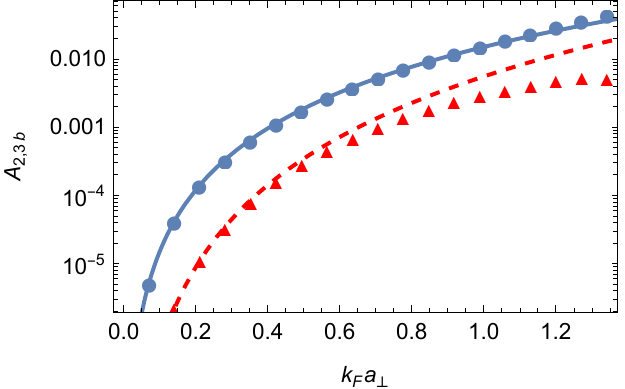}
    \caption{Contributions of 2 and 3-body effective interactions to the first beyond-mean-field corrections. Blue circles: 2 body contribution, Blue solid line: asymptotic expression (\ref{Eq:A2b}). Red triangle: three body contribution $A_{3b}$. Red dashed line: asymptotic expression (\ref{Eq:A3b}).}
    \label{fig:A23b}
\end{figure}

Since the corrections are proportional to $g_{3D}^2$, they will scale as $\gamma^2$. We can write the full energy as 

\be
E=E_{\rm YG}(\tilde g_{1D})\left[1-(A_{\rm 2b}+A_{\rm 3b})\gamma^2
\right],
\ee
where the first term comes from finite range corrections and the second one from three-body interactions. The sum appearing in Eq. (\ref{Eq:gdstate2}) can be calculated analytically in the quasi-1D limit  $E_F\ll \hbar\omega_\perp$ (see appendix \ref{Sec:A2}) and we have 
\begin{eqnarray}
A_{2\rm b}&\simeq&\frac{\zeta(3/2)}{2\pi^3}\left(\frac{E_F}{\hbar\omega_\perp}\right)^{3/2}\label{Eq:A2b}\\
A_{3\rm b}&\simeq&\frac{8}{\pi^4}{\rm Li}_{2}(1/4)\left(\frac{E_F}{\hbar\omega_\perp}\right)^2\label{Eq:A3b}
\end{eqnarray}
where $\zeta$ is Riemann's zeta function and where ${\rm Li}_s(z)$ is a polylog function as before . These asymptotic behaviours are compared to numerical calculations in Fig. \ref{fig:A23b}.  The approximate result provides an accurate value for $A_{\rm 3b, 2b}$, even for $E_F\simeq \hbar\omega_\perp$ in the case of two-body interactions. We also note that the two body-contribution always dominate its three-body counterpart. Finally, even when for $E_F\simeq \hbar\omega_\perp$, both ${A_{2b}}$ and $A_{3b}$ do not exceed $\simeq 5\times 10^{-2}$, which suggests that virtual transitions affect only weakly the first beyond-mean-field corrections to the energy.

\section{Discussion}
In this article we have calculated derived the effective many-body Hamiltonian describing a quasi-1D cloud of spin 1/2 fermions. We have calculated the first beyond mean-field corrections to the energy of a cloud of spin 1/2 fermions confined in a quantum wire. Our work extend to the fermionic many-body regime previous results on the few-body bosonic problem in quasi 1D. We have shown that, even for Fermi energies close to $\hbar\omega_\perp$ the corrections due to virtual transitions towards transverse excited states are rather small up to $\gamma\simeq 1$ where our calculation provides an accurate estimate for the energy. This means that experiments using cold atoms in quantum wires provide an accurate description of the ground state of Yang-Gaudin's Hamiltonian  that the most important source of discrepancy with pure one-dimensional physics will most likely be the occupation of transverse excited states due to the finite temperature of the system. It would then be interesting to connect our work to high temperatures results obtained using Virial expansion \cite{Kristensen2016Second-orderWaveguide}.  

Even though the Yang-Gaudin provides a good description of the weakly-interacting regime, we have shown that it fails in the strongly attractive limit. In future work we will therefore extend our calculation to the non-perturbative limit to explore the breakdown of the 1D effective regime. Another intriguing research direction would be to generalize our results to quasi-2D systems where puzzling results showing deviations between experiments and 2D theories have been reported in \cite{Sobirey2021ComparingDimensions}. 
\acknowledgements

We thank Antoine Browaeys,  Christophe Salomon, Lovro Barisic and Xavier Leyronas for insightful discussions. This work was supported by R\'egion Ile de France (DIM Sirteq, project 1DFG) and CNRS (grant 80Prime and Tremplin). 

\appendix

\section{Derivation of the effective Hamiltonian}
\label{Appendix1}
%In order to determine the function $f$, we substitute %Eq.(\ref{SwS}) into Eq.(\ref{SWa}) and calculate the 
%left-hand side based on the anticommutation relations %$\{a_{\sigma k m}^\dagger, a_{\sigma^\prime p %n}\}=\delta_{\sigma\sigma^\prime} \delta_{kp}\delta_{mn}$, %$\{a_{\sigma k m}, a_{\sigma^\prime p n}\}=0$
%together with the identities $[AB,C]=[A,C]B+A[B,C]$ and $[AB,C]=A\{B,C\}-\{A,C\}B$. Comparing the result with Eq.(\ref{H2}), we obtain:
%\begin{equation}\label{f}
%f(\{k_i,m_i\})= \frac{\chi_{m_1 m_2 m_3 m_4}\delta_{k_1+k_3,k_2+k_4}}{\epsilon(k_1,m_1)+\epsilon(k_3,m_3)-
%\epsilon(k_2,m_2)-\epsilon(k_4,m_4)}.   
%\end{equation}

We provide below the second quantization expression of the
effective Hamiltonian $H_{\rm eff}$ obtained via the SW transformation. To this end, we substitute the explicit form of the generator $S$, given in Eq.s (\ref{SwS}) and (\ref{f}), into Eq.(\ref{Heff}) and evaluate the two commutators $[S,H_1]$ and $[S,H_2]$. The action of each commutator can be written as a sum of two parts: the first part contains all contributions with products of four field operators and therefore renormalizes two-body interactions, while the second part  gathers all contributions with products of six field operators, thus describing  effective three-body interactions. For the  commutator involving $H_2$, an explicit  calculation yields
\begin{widetext}
\begin{equation}
\begin{split}
&\frac{1}{2}[S,H_2]   = \frac{g_{3D}^2}{2L^2} \Bigg[ \;\sum_{\substack{k_1+k_3=p_2+p_4\\\    \alpha_1,\alpha_3,\beta_2,\beta_4}} 
     \left [ F_{\alpha_1 \alpha_3}^{ \beta_2 \beta_4}(k_1,k_3)+\bar F_{\beta_2 \beta_4}^{\alpha_1 \alpha_3}(p_2,p_4)\right ]\;
    a_{\uparrow k_1 \alpha_1}^\dagger  a_{\downarrow k_3 \alpha_3}^\dagger a_{\downarrow p_4 \beta_4} a_{\uparrow p_2 \beta_2} \\
&+ \sum_{\substack{k_1+k_3+p_3=p_2+p_4+k_4\\\alpha_1,\alpha_3,\alpha_4,\beta_2,
\beta_3,\beta_4}} 
 \!\!\!\left [G_{\alpha_1 \alpha_3 \alpha_4}^{\beta_2 \beta_3 \beta_4} (k_1, k_3,k_4)+\bar G_{\beta_2 \beta_4 \beta_3}^{\alpha_1 \alpha_3 \alpha_4}(p_2, p_4, p_3) \right ] \left (a_{\uparrow k_1 \alpha_1}^\dagger   a_{\downarrow k_3 \alpha_3}^\dagger a_{\downarrow p_3 \beta_3 }^\dagger a_{\downarrow p_4 \beta_4} a_{\downarrow k_4 \alpha_4} a_{\uparrow p_2 \alpha_2} + (\uparrow \leftrightarrow \downarrow) \right ) \Bigg],
\end{split}
\label{contr1}
\end{equation}
\end{widetext}
where the overbar stands for complex conjugation and  
we have introduced the functions
%
%\begin{eqnarray}
\be
F_{\alpha_1 \alpha_3}^{\beta_2 \beta_4}(k_1,k_3)=\sum_{k_2 k_4 \alpha_2 \alpha_4} 
f(\{k_i,\alpha_i\}) \chi_{\alpha_2 \beta_2 \alpha_4 \beta_4}\;\;\; \label{defF}\\
\ee

\be
G_{\alpha_1 \alpha_3 \alpha_4}^{ \beta_2 \beta_3 \beta_4} (k_1, k_3,k_4)=\sum_{k_2 \alpha_2}    f(\{k_i,\alpha_i\}) \chi_{\alpha_2 \beta_2 \beta_3 \beta_4}. \label{defG}
\ee
%\end{eqnarray}
%\null
%\newpage
A similar calculation for the commutator $[S,H_1]$ 
%last term in the rhs of Eq.(\ref{Heff}), 
gives
\begin{widetext}
\begin{eqnarray}
[S,H_1] &  =& \frac{g_{3D}^2}{L^2}  \sum_{\substack{k_1+k_3=p_2+p_4\\(m_1,m_3) \not =(0,0)}} 
    \Big[ R_{\alpha_1 \alpha_3}(k_1,k_3) a_{\uparrow k_1 \alpha_1}^\dagger a_{\downarrow k_3 \alpha_3}^\dagger a_{\downarrow p_4 0} a_{\uparrow p_2 0} + \bar R_{\alpha_1 \alpha_3}(k_1,k_3)
     a_{\uparrow p_2 0}^\dagger  a_{\downarrow p_4 0}^\dagger a_{\downarrow k_3 \alpha_3}
     a_{\uparrow k_1 \alpha_1} \Big]
 \nonumber\\
&+&\frac{g_{3D}^2}{L^2} \sum_{\substack{k_1+k_3+p_3=p_2+p_4+k_4\\ (\alpha_1,\alpha_3,\alpha_4) \not =(0,0,0)}} \Bigg[
 T_{\alpha_1 \alpha_3 \alpha_4} (k_1, k_3,k_4) \left (
%a_{\uparrow k_1 \alpha_1}^\dagger   a_{\downarrow k_3 \alpha_3}^\dagger a_{\downarrow p_3 0}^\dagger a_{\downarrow p_4 0} a_{\downarrow k_4 \alpha_4} a_{\uparrow p_2 0} + (\uparrow \leftrightarrow \downarrow)\right )  \nonumber\\
%&+&  \bar T_{\alpha_1 \alpha_3 \alpha_4} (k_1, k_3,k_4) \left ( a_{\uparrow p_2 0}^\dagger 
a_{\downarrow k_4 \alpha_4}^\dagger a_{\downarrow p_4 0}^\dagger a_{\downarrow p_3 0} a_{\downarrow k_3 \alpha_3} a_{\uparrow k_1 \alpha_1} + (\uparrow \leftrightarrow \downarrow)\right ) \Bigg],
   \label{contr2}
\end{eqnarray}
\end{widetext}
where we have introduced the functions
\begin{eqnarray}
T_{\alpha_1 \alpha_3 \alpha_4}(k_1,k_3,k_4)=\sum_{k_2} \chi_{\alpha_1 0 \alpha_3 \alpha_4}\chi_{0 0 0 0} \delta_{k_1+k_3,k_2+k_4}\nonumber \\
\! \!\! \times  \frac{1}{\epsilon(k_1,\alpha_1)+\epsilon(k_3,\alpha_3)-
\epsilon(k_2,0)-\epsilon(k_4,\alpha_4)} \quad \label{T}
\end{eqnarray}
and
\begin{equation}\label{R}
 R_{\alpha_1 \alpha_3}(k_1,k_3)=\sum_{k_4} T_{\alpha_1 \alpha_3 0}(k_1,k_3,k_4).  %\delta_{k_1+k_3,k_2+k_4} \\
\end{equation}
Notice that the Hamiltonian correction in Eq.(\ref{contr2}) does not contribute to the ground state energy of the system, because it does not contain terms with all $\alpha_i=0$.

\section{Calculation of the sums appearing in the two- and three-body effective interactions }
\label{Sec:A2}

When calculating the energy of the quasi-1D gas in Eq. (\ref{Eq:gdstate2}), two types of sums appear. Firstly, we have terms involving $\chi_{0\alpha 00}$. This matrix element can be explicitly written as 

$$
\chi_{0\alpha 00}=\int d^2\bm\rho \psi_0 (\bm \rho)^3 \psi_\alpha (\bm\rho). 
$$
where we have used the fact that the ground-state wavefunction $\psi_0$ is real.
Since the ground state of the 2D harmonic oscillator is isotropic we first see that the function $\psi_\alpha$ must correspond to a null angular momentum. In this case $n_\alpha$ is necessarily even and we can then write, using the general properties of the 2D harmonic oscillator
$$
\psi_\alpha(\rho)=\frac{1}{\sqrt{\pi}a_\perp}L_{n_\alpha/2}(\rho^2/a_\perp^2)e^{-\rho^2/2a_\perp^2},
$$
where $L_n(x)$ is the Laguerre polynomial of order $n$. 

Taking $u=\rho^2/a_\perp^2$, we have thus
$$
\chi_{0\alpha 00}=\frac{1}{\pi a_\perp^2}\int du L_{n_\alpha/2} (u) e^{-2 u}=\frac{1}{2\pi a_\perp^2}\frac{1}{2^{n_\alpha/2}}, 
$$
where we have used the value of the Laplace transform of a Laguerre polynomial
$$
\int_0^\infty du e^{-s u}L_n(u)=\frac{1}{s}\left(\frac{s-1}{s}\right)^n.
$$
Thanks to the exponential decay $\chi_{0\alpha 00}$, the associated sum over $\alpha$ appearing in Eq. (\ref{Eq:gdstate2}) converges.

The second type of sum has the general structure
\begin{eqnarray}
A(z)&=&\sum_{\alpha_1\alpha_2}\frac{|\chi_{0\alpha_1 0\alpha_2}|^2}{z-\hbar\omega_\perp (n_{\alpha_1}+n_{\alpha_2})}\\
&=&\sum_{\alpha_1\alpha_2}\frac{|\langle\alpha_1\alpha_2|\delta (\bm\rho_{12})|00\rangle|^2}{z-\hbar\omega_\perp (n_{\alpha_1}+n_{\alpha_2})}.
\end{eqnarray}
with $\bm \rho_{12}=\bm\rho_1-\bm\rho_2$. Noting that the sum is actually a completeness relation, this expression can be recast as 
$$
\langle 00|\delta (\bm\rho_{12})\frac{1}{z+2\hbar\omega_\perp-h_1-h_2}\delta(\bm\rho_{12})|00\rangle,
$$
where $h_i$ is the transverse harmonic oscillator Hamiltonian describing the dynamics of particle $i$ in the $(x,y)$ plane, up to a constant to set the ground state energy at 0. 
Instead of describing the state of the system using the coordinates of the two particles, we can consider the relative and center of mass degrees of freedom of the pair of atoms. In this case the Hamitonian $h_1+h_2$ can be written as a sum of two harmonic oscillator $h_r$ and $h_c$ associated respectively with the relation and center of mass motion of the pair. They correspond to harmonic oscillators of the same frequency $\omega_\perp$ and masses $\mu=m/2$ and $M=2m$. If we insert a completeness relation for the new basis $|\alpha_r,\alpha_c\rangle$, we note that since $\delta(\bm\rho_{12})$ acts only on the relative motion, we do not have any contirbution of the center of mass degrees of freedom and the  sum can be written as

$$
A(z)=\sum_{\alpha_r}\frac{|\langle\alpha_r|\delta (\bm\rho_{12})|0\rangle|^2}{z-\hbar\omega_\perp n_{\alpha_r}}=\sum_{\alpha_r}\frac{|\psi_{\alpha_r}(0)\psi_0(0)|^2}{z-\hbar\omega_\perp n_{\alpha_r}}
$$
Since the sum depends only on the value of the wavefunction in $\bm\rho_{12}=0$, it means that only the zero-angular momentum modes contribute. In this case, $n_{\alpha_r}$ is even and $\psi_{\alpha_r}(0)=1/\sqrt{2\pi}a_\perp$. We finally have
$$
A(z)=\frac{1}{(2\pi)^2 a_\perp^4}\sum_{n=0}^\infty\frac{1}{z-2\hbar\omega_\perp n}
$$

Strictly speaking, this sum is divergent. However, in Eq. (\ref{Eq:gdstate2}), we take the difference of two terms having the same structure and compensating at large $n$. Note also that in that equation, the sum over $(\alpha_1,\alpha_2)$ excludes the ground state, which amounts to starting the sum over $n$ at $n=1$.

\bibliography{references.bib}

\end{document}